%% file: main.tex
\documentclass[lettersize,journal]{IEEEtran}
\usepackage{amsmath,amsfonts}
\usepackage{algorithmic}
\usepackage{algorithm}
\usepackage{array}
\usepackage[caption=false,font=normalsize,labelfont=sf,textfont=sf]{subfig}
\usepackage{textcomp}
\usepackage{stfloats}
\usepackage{url}
\usepackage{verbatim}
\usepackage{graphicx}
\usepackage{cite}
\usepackage{xcolor}
\usepackage[originalparameters]{ragged2e}
\newcolumntype{x}[1]{>{\RaggedRight}p{#1}}

\definecolor{brickred}{rgb}{0.8, 0.25, 0.33}

\usepackage{booktabs}
\usepackage{multirow}
\newcommand{\ra}[1]{\renewcommand{\arraystretch}{#1}}

\usepackage{array}
\newcolumntype{L}[1]{>{\raggedright\let\newline\\\arraybackslash\hspace{0pt}}m{#1}}
\newcolumntype{C}[1]{>{\centering\let\newline\\\arraybackslash\hspace{0pt}}m{#1}}
\newcolumntype{R}[1]{>{\raggedleft\let\newline\\\arraybackslash\hspace{0pt}}m{#1}}

\makeatletter
 \let\old@ps@headings\ps@headings
 \let\old@ps@IEEEtitlepagestyle\ps@IEEEtitlepagestyle
 \def\confheader#1{%
 \def\ps@IEEEtitlepagestyle{%
 \old@ps@IEEEtitlepagestyle%
 \def\@oddhead{\strut\hfill#1\hfill\strut}%
 \def\@evenhead{\strut\hfill#1\hfill\strut}%
 }%
 \ps@headings%
 }
\makeatother

\confheader{%
\parbox{18cm}{\centering Accepted copy for Publication at IEEE Journal of Radio Frequency Identification, vol. 6, pp. 637-648, 2022 \\ 
Final published version available at: https://doi.org/10.1109/JRFID.2022.3170381}}

\begin{document}

\title{Secure and Trustworthy NFC-based Sensor Readout for Battery Packs in Battery Management Systems}

\author{
\IEEEauthorblockN{Fikret Basic, Martin Gaertner, Christian Steger} \\
\IEEEauthorblockA{\textit{Institute for Technical Informatics} \\
\textit{Graz University of Technology}\\
Graz, Austria \\
\{basic, steger\}@tugraz.at}
}

\maketitle

\begin{abstract}
Wireless Battery Management Systems (BMS) are increasingly being considered for modern applications. The ever-increasing complexity and production costs of BMS modules and wired connections resulted in a necessity for new ideas and approaches. Despite this growing trend, there is a lack of generic solutions focused on battery cells' sensor readout, where wireless communication allows for a more flexible and cost-efficient sensor installation in battery packs. Many wireless technologies, such as those that use the 2.4\,GHz frequency band, suffer from interference and other limitations. In this article, we present an alternative approach to communication in BMS that relies on the use of Near Field Communication (NFC) technology for battery sensor readouts. As an answer to the rising concern over the counterfeited battery packs, we consider an authentication schema for battery pack validation. We further consider security measures for the processed and stored BMS status data. To show that a general BMS application can make use of our design, we implement a BMS demonstrator using the targeted components. We further test the demonstrator on the technical and functional level, by also performing evaluation on its performance, energy usage, and a security threat model.
\end{abstract}


\section{Introduction}
\label{sec:intro}
\input{introduction}

\section{Background and Related Work}
\label{sec:rel_work}
\input{related_work}

\begin{figure}[!ht]
  \centering
  \includegraphics[width=0.95\linewidth]{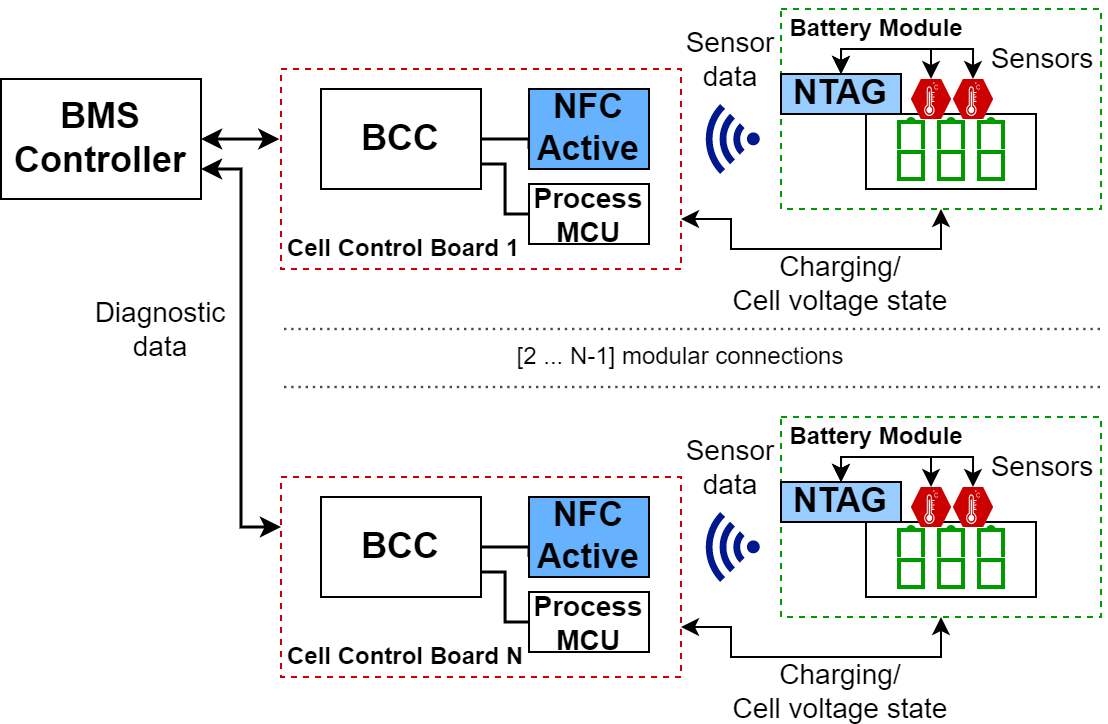}
  \caption{Proposed BMS modular design architecture utilizing NFC components.}
  \label{fig:bms_design_architecture}
\end{figure}

\section{Design of the Novel BMS NFC Sensor Readout}
\label{sec:design}
\input{design}

\section{Security Mechanisms}
\label{sec:sec_mech}
\input{security_mech}

\section{Evaluation}
\label{sec:evaluation}
\input{evaluation}

\section{Discussion and Future Work}
\label{sec:discussion}
\input{discussion_future_work}

\section{Conclusion}
\label{sec:conclusion}
\input{conclusion}

\section*{Acknowledgments}
We would like to express our sincere thanks to the BMS team from the NXP Semiconductors Austria GmbH Co \& KG for the support and cooperation during the design and evaluation phases of this work, as well as for providing the necessary equipment for conducting our experimental tests.

This project has received funding from the ``EFREtop: Securely Applied Machine Learning - Battery Management Systems'' (Acronym ``SEAMAL BMS'', FFG Nr. 880564).

\bibliographystyle{ieeetr}
\bibliography{references}


\begin{IEEEbiography}[{\includegraphics[width=1in,height=1.25in,clip,keepaspectratio]{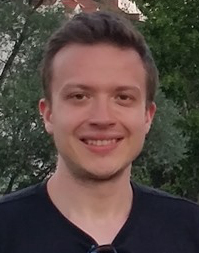}}]{Fikret Basic}
received the Dipl.-Ing. (M.Sc.) degree in computer science 
from the Graz University of Technology in 2019, with the main focus of his studies being on the pervasive computing, information systems, secure system design and HW/SW codesign. From 2019 until 2020 he worked as a design engineer in CISC Semiconductors. From 2020 he is employed at the Institute of Technical Informatics of Graz University of Technology, where he is currently also pursuing the Ph.D. degree in information and computer engineering. His current area of research focuses on the security in battery management systems.
\end{IEEEbiography}

\begin{IEEEbiography}[{\includegraphics[width=1in,height=1.25in,clip,keepaspectratio]{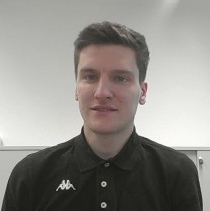}}]{Martin Gaertner}
received the Dipl.-Ing. (M.Sc.) degree in information and computer engineering
from the Graz University of Technology in 2021, focussing on secure and correct systems and embedded automotive
systems. His research focuses on secure systems in embedded automotive environments.
\end{IEEEbiography}

\begin{IEEEbiography}[{\includegraphics[width=1in,height=1.25in,clip,keepaspectratio]{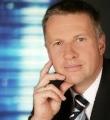}}]{Christian Steger}
received the Dipl.-Ing. (M.Sc.) degree in 1990, and the Dr.Techn. (Ph.D.) degree in electrical engineering from the Graz University of Technology, Austria, in 1995. He graduated from the Export, International Management and Marketing course with the Karl-Franzens-University of Graz in 1993 and completed the Entrepreneurship Development Program at MIT Sloan School of Management, Boston, in 2010. He is a Strategy Board Member of the Virtual Vehicle Competence Center (ViF, COMET K2) in Graz, Austria. Since 1992, he has been an Assistant Professor with the Institute of Technical Informatics, Graz University of Technology, where he heads the HW/SW codesign group with the Institute of Technical Informatics.
\end{IEEEbiography}

\end{document}

%% file: introduction.tex
\IEEEPARstart{G}{reen} energy and sustainability are becoming more important than ever before, with Battery Management Systems (BMS) also seeing an increased interest from the industry and the research community. This resulted in a higher digitization and use-case expansion that required additional attention to the already complex systems that utilize BMS~\cite{Hu2019}.
BMS play an important role in many systems today that rely on the use of large battery packs. They are often mentioned as one of the main critical controller components being part of smart power grids and electric or hybrid vehicles~\cite{8168251, 6532486}. They are used as control devices in such systems, where they regulate the usage of individual battery cells by offering monitoring and diagnostic services, as well as the possibility to track the lifetime usage of each individual cell \cite{andrea2010battery}. To offer safer and more efficient energy usage, a BMS also handles cell balancing control during the charging and discharging cycles.
A BMS can be deployed in different topologies and usually consists of various devices. They generally contain a central BMS controller, which in a modulated setting, communicates with individual Battery Cell Controllers (BCCs). The BCCs help in relaying diagnostic data back to the BMS controller through monitoring and control of individual battery cells. These cells are packed together in parallel or serial connections inside battery modules, with accompanying temperature, pressure, or other sensors. Data received from these sensors is critical in preventing dangerous incidents like the thermal runaway, which happens due to the rapid increase in battery temperature~\cite{fireBattery}. A BMS controller is able to derive diagnostic data based on the monitored and measured data from the battery cells, e.g., State of Charge (SoC), State of Health (SoH), etc., \cite{Hu2019, 8168251, 6532486}.  
However, these controllers can only act as long as they have the correct information on the current state inside the battery pack, i.e., they are dependent on the sensor readouts. Two main factors influence the accuracy of these readings: (i) the number of sensors used, and (ii) the relative position of a sensor to its measurement target. 

Traditionally, BMS use wired connection to handle the communication between individual modules, i.e., between the BCCs and the battery cell sensors. 
This, however, imposes several limitations, as shown in Table~\ref{table:wired_bms_limits}. 
In order to alleviate the aforementioned wired limitations, it is possible to replace wired with wireless technology networks. We see several challenges that need to addressed when choosing an appropriate wireless technology, as indicated in Table~\ref{table:chg_wireless_bms}.

Different communications have already been tested in an attempt to solve the mentioned limitations. Research with Bluetooth~\cite{7357002, 7151581} and ZigBee~\cite{Rahman_2017} have been tested and evaluated within the BMS domain. However, while they give promising results for the data throughput, these studies fail to address the main challenge of the mentioned technologies, that being the interference. Security is also only partially covered, mostly under the given technologies' security stack, with ZigBee being especially subjected to limited throughput and security concerns \cite{electronics10182193}. Schneider et al.~\cite{6229439} address most of the concerned challenges but does not focus on the security aspects and newer modulated BMS considerations. 
We further discuss the BMS wireless and security findings, and their relevance to our work, in Section~\ref{sec:rel_work}.

\begin{table}[!ht]
\ra{1.1}
\caption{Limitations of the Wired BMS~\cite{electronics10182193, 7385582, 6914889, wireless_battery_arpa, wired_vs_wireless}} 
\label{table:wired_bms_limits}
\begin{center}
\begin{tabular}{@{}lm{6.5cm}@{}}
\toprule
    Limitation & \multicolumn{1}{C{6.5cm}}{Description}  \\ \midrule
    \textit{Assembly cost} & Each connection to and from the cell and sensor source needs to be physically soldered, and it needs to account for materials used. \\
    \midrule[0.15pt]
    \textit{Scalability} & The complexity of design, deployment, and afterwards maintenance, boosts with the increase in the number of battery cell sensors. \\
    \midrule[0.15pt]
    \textit{Area coverage} & Due to the physical wires used, sensor placement is often limited. This can make the use of certain areas impossible, e.g., inner housing of the cells. \\
\bottomrule
\end{tabular}
\end{center}
\end{table}

\begin{table}[!ht]
\ra{1.1}
\caption{ Wireless Technology Challenges for BMS}
\label{table:chg_wireless_bms}
\begin{center}
\begin{tabular}{@{}m{1.4cm}m{6.8cm}@{}}
\toprule
    Challenge & \multicolumn{1}{C{6.8cm}}{Description}  \\ \midrule
    \RaggedRight \textit{Restricted throughput} & For a safe and continuous execution of BMS operations, it is necessary to maintain a steady and fast flow of data from the battery cell sensors to the BCCs and afterwards to the BMS controller \cite{electronics10182193, wired_vs_wireless}. \\
    \midrule[0.15pt]
    \textit{Interference} & The use of the same frequency band, even through different wireless technologies, can cause interference. This is especially true with the 2.4 GHz band, which is used by several technologies, most prominent being LR-WPAN, Bluetooth, ZigBee, and WiFi\cite{8591253, 7357002, 7151581, Rahman_2017}. \\
    \midrule[0.15pt]
    \RaggedRight \textit{Multipath propagation} & System's resilience in
    maintaining communication sight and reliability under obstructive environments \cite{6966212}. \\
    \midrule[0.15pt]
    \RaggedRight \textit{Security concerns} & Unless placed in an enclosed case, wireless networks are prone to eavesdropping, remote attacks, and other malicious incursions~\cite{9090905, Sripad2017, Kumbhar2018}. \\
\bottomrule
\end{tabular}
\end{center}
\end{table}

To address the BMS wired limitations and wireless requirements, we propose a system architecture for modular BMS that offers the NFC technology for battery module cells' sensor readout. This includes the extension with the conventional BCC by adding a connection to an active NFC reader. The battery cell's sensors would, hence, only connect to the provided passive NFC device per battery pack module, not requiring additional connection or power draw for their functionality. Security and data processing are handled via an additional Microcontroller Unit (MCU). These additional components would form a new overall control block together with the BCC. This block would still remain modulated, i.e., it would maintain the same input and output connections. A BMS with our presented architecture is able to perform security operations on the logged status data with a minimal overhead increase, while retaining its original functionality.

\noindent\textbf{Contributions:} In this work, we present an answer to the listed challenges, by proposing a design model that utilizes NFC as the chosen technology for the wireless communication between the battery cell sensors and the BCCs. By introducing NFC, we are not only able to answer to the design restrictions imposed through the use of wired communication, but also address the challenges introduced when using wireless communication technologies.

After providing the relevant background information and presenting related work in Section~\ref{sec:rel_work}, we make the following contributions in this article:
\begin{itemize}
    \item We present a novel approach for the NFC-based BMS battery cells' sensor readout by indicating design points for the device use and placement, as well as the exchange protocol between the modules (Section~\ref{sec:design}). 
    \item We address the battery cell source validity \cite{8813669, 101093} question by proposing an authentication model for verifying individual battery cell modules. BCCs should communicate only with the trusted battery cells, for the reasons of both security and safety concerns (Section~\ref{sec:secure_mech_auth}).
    \item We investigate the security protocol and design requirements for the purpose of storing and securely handling the derived BMS status data as the next operational step after the sensor readout (Section~\ref{sec:secure_mech_diag_prot} \& \ref{sec:secure_mech_sec_proc}). 
    \item We experimentally show the feasibility of our design by realizing a BMS system prototype and implementing NFC and security control functionalities. The system is further evaluated on its (i) security dependability by a threat model analysis, (ii) time measurements for individual BMS NFC readout phases, (iii) protocol overhead analysis for the secure BMS monitoring and diagnostic data logging (Section~\ref{sec:evaluation}), (iv) system energy consumption, and (v) potential NFC sensor readout throughput.
\end{itemize}

This article presents an extended version of the published paper~\cite{9617320} that includes a more detailed analysis and investigation of the proposed design specifications related to the NFC integration for the BMS inter-module communication and sensor data readout. On top of the design and authentication approach presented in that paper, an additional security investigation and evaluation for the purpose of securely logging sensor monitoring and diagnostic BMS data were conducted. Additionally, in relation to the NFC system design, a throughput and energy consumption analysis have been done as well.

%% file: related_work.tex
\subsection{Wireless Battery Management Systems (WBMS)}

The increase of the number of battery cells in modern BMS resulted in an increase in the number of used component devices, especially regarding intermediate control components. This all further lead to an increase in expenses and complexity in cable installation. New topologies and architectures had to be introduced focused on using wireless technologies. Primarily, they were seen as an extension to already different derivations of modular and distributed BMS~\cite{andrea2010battery}. Some of the pioneering research includes the realization of a WBMS under custom chips and protocols by M. Lee et al.~\cite{6914889}. In this work, they introduce a WiBaAN protocol that works under the 900\,MHz band with a data rate of up to 1\,Mbit/s, allowing for direct communication between a large set of battery cells and the main BMS controller. However, while novel for the time of publishing, the relatively low data throughput rate, used frequency band, manufacturing costs, and no newer research updates regarding the modulated BMS topologies could present a limitation for the modern BMS derivations.

Several design models have been proposed and investigated in the domain of the 2.4 GHz frequency band.
Shell et al.~\cite{7151581} presents a Bluetooth-based BMS design approach. They show its feasibility under the standard BMS environment and commercial applications. De Maso-Gentile et al.~\cite{7357002} presents a different design approach, that is more focused on applying Bluetooth gateway access to already conventional BMS CAN infrastructures. However, most of the proposals based on Bluetooth technology are primarily centred on intra-module communication and do not account for direct battery sensor readout. Bluetooth, specifically the newer BLE, has a limited throughput rate which can often fluctuate due to noisy channels even in the newer 5.x standards \cite{10.1145/3331052.3332471}. This can make it difficult to fulfill the necessary standard requirements for data transfer under the conventional BLE topologies.  
Research has been also conducted using the ZigBee technology by Rahman et al.~\cite{Rahman_2017}. While it showed potential in its applicability, ZigBee would suffer from restrictions due to its low-data rates and unstable channels. 
Wi-Fi was also considered under specialized BMS investigations. Gherman et al.~\cite{8591253} propose a WBMS build on a single chip that used Wi-Fi as its communication technology for their demonstrator. A different kind of research, more focused on smart cells, was proposed by Huang et al.~\cite{9236279}. Here, the communication between the individual cells and the main controller is done over a Wi-Fi channel, with the BMS controller using the channel for the cell balancing control. The focus of this research was on cell balancing and smart cells, with Wi-Fi being mostly used as a demonstrative wireless technology with no significant focus on the wireless aspects and challenges.
The presented research gives an insight into the communication between the BCCs or similar modules and the main BMS controller using the prescribed wireless technologies.
Also, as mentioned in Section~\ref{sec:intro}, 2.4\,GHz technologies generally suffer from an increased chance of interference under complex environments, e.g., Electric Vehicles (EVs), where many devices and modules could compete over the use of the bandwidth channels.
This work extends the wireless usage regarding the BMS components by also focusing on the BCC and sensor communication, which is often overlooked, through NFC utilization. 

BMS today are also often considered for the cloud service extensions \cite{en11010125, LI2020101557}. These solutions offer the distribution of BMS modules over a wider area, and hence, further reduce the use of wires and deployment complexity. They also provide functionality extensions. Cloud services aim to cover the calculation of important State of Health (SoH) and State of Charge (SoC) BMS functions on a more efficient cloud base, by using different data sources and even resource-demanding machine learning algorithms, which otherwise would not be possible on resource and process constrained BMS MCU field controllers. 
These services, however, are outside of the scope of this work, as they focus mainly on the external, rather than on the internal BMS communication.

\subsection{Security in Battery Management Systems (BMS)}

The current research work related to BMS security is limited, due to it being a relatively novel topic that first started sparking interest in the recent time. Nonetheless, there has already been some research done focused on different aspects of the BMS security design. Sripad et al.~\cite{Sripad2017} present an investigation of the cybersecurity threats of BMS, particularly of EVs, especially related to their interaction with battery packs and to overcharging and discharging manipulation concerns. A FACTS approach proposed by Khalid et al. in \cite{8813669} deals with a formal threat analysis of BMS by investigating and comparing different existing frameworks. It also goes into a detailed analysis, points out and classifies important general security threats found under a BMS. Further BMS threat analysis models have also been proposed by Kumbhar et al.~\cite{Kumbhar2018}. This work also goes in a direction of a wider topic and includes some security overview of BMS Internet-of-Things (IoT) solutions. A similar work that looks at the IoT security perspective with BMS and their related environments is by Lopez et al.~\cite{Lopez2017}. While most of the mentioned publications present a broad BMS security analysis topic, they still serve as a good starting ground to complement the presented work. 

\subsection{Near Field Communication (NFC) Applications}

NFC is a high frequency (HF) communication standard based on the Radio Frequency Identification (RFID) which operates on a frequency band of 13.56\,MHz, has a typical range of up to $10\,cm$, and depending on the standard, supports data rates of up to 848\,kbit/s \cite{ISO18092, ISO21481}. It handles different modes of communication, among them being communication between an active reader and a passive tag device. Like the RFID, it supports the energy harvesting features from the active to the passive device during the data exchange.
The use of the NFC technology in more extensive system infrastructures has already been investigated before. Specifically, research presented by Ulz et al. \cite{8098906} proposes the use of NFC-based communication for robot-machine interaction in an Industry 4.0 setting. Additionally, work by Chen et al.~\cite{6228335} investigates secure authentication and anti-counterfeiting methods using RFID. Alzahrani et al.~\cite{9075232} propose an NFC-focused anti-counterfeiting system. Despite a large amount of research being done both for the general wireless BMS and the integration of NFC in similar environments, not much specific work has yet been done that combines these two fields of interest, which is also indicated by the recent survey research paper by A. Samanta and S. S. Williamson~\cite{electronics10182193}. Work done by Schneider et al.~\cite{6229439} focuses largely on this field by also proposing a design approach for wireless BMS battery sensors utilizing the same RFID technology. However, one of the main focal points in that paper is placed on the issues caused by galvanic isolation. Moreover, due to the date when the paper was published, it does not account for the newer BMS modular architectures and modern NFC derivations, alongside the security aspects. In this work, we try to bridge that gap and show the potential of using NFC in hard-to-reach sensor environments while at the same time giving attention to the security requirements.

%% file: design.tex
For the targeted design architecture we divide the entire system into three main modules:
\begin{itemize}
    \item \textit{BMS controller}
    \item \textit{Cell control board} (CCB)
    \item \textit{Battery module}
\end{itemize}

Modules, as well as their placement and connections, are illustrated in Fig.~\ref{fig:bms_design_architecture}. Here, the BMS controller plays the role of the main control unit responsible for receiving and interpreting diagnostic data and conducting necessary safety control actions. It can contain one or multiple operational MCUs. The BMS controller communicates with the CCB that contains a BCC, an NFC reader as the communication interface, and optionally, an additional control MCU for the protocol handling connected via a supported communication bus protocol, e.g., Serial Peripheral Interface (SPI). In a traditional design, the CCB usually only contains a simple BCC aimed primarily only for the BMS functional support. To supplement the communication handling requirements for the NFC reader, and also preceding and subsequent data and security processing, we introduce another process MCU that works either along with the BCC chip or on top of the BCC functional block. In most scenarios, the communication between the CCBs and the BMS controller is established using either a Controller Area Network (CAN) protocol, or some other form of network connection, like Ethernet, Transformer Physical Layer (TPL), or SPI \cite{cyber_sec_book}.One BMS controller can communicate with multiple CCBs, depending on the system and protocol limitations \cite{Hu2019, andrea2010battery}.

The battery module contains battery cells, sensors, and an NFC communication interface to the CCB. In the presented design, this interface is an \textit{NFC-Tag} (NTAG). The communication for the NTAG and sensors is primarily done with the Inter-Integrated Circuit (I2C) protocol.
For charging and discharging cycles, as well as related voltage readings, we still rely on the hardwired measurements from conventional modulated BMS designs, with them being usually less demanding in terms of placement and installation compared to the investigated sensor connections and also requiring an otherwise special handling. 

\subsection{NFC Communication}

To make the communication between the BCCs and battery cells using the wireless NFC technology possible, appropriate devices and communication modes need to be chosen. In the presented design, \textit{Reader/Writer mode} is opted as the chosen mode of communication. The NFC reader plays the role of the active device that is connected to a specialized controller BCC, as well as to an MCU for pre-processing and security operations. In traditional designs of the modulated BMS, this MCU can also be already found as an integral part of the BCC. It is, however, vital, that the main functionality conditions are fulfilled and contained which entail that the communication over the NFC can be accurately processed and handled, as well as to be able to handle security operations. Before the communication begins, the NFC reader needs to have discovered the targeted NTAG(s) using the discovery loop process. Immediately afterwards, the authentication process starts. It is important to make sure that no formal communication can begin before the battery modules have been authenticated, as to avoid any potential vulnerabilities that might arise afterwards. Following the successful authentication, the NTAG proceeds to initiate self-configuration and prepares to communicate with both the sensors and the NFC reader. Since in a standard environment, the same devices are going to be also used for the subsequent measurement readings, the initialization and configuration steps can be cached and therefore omitted. 

\begin{figure*}[!ht]
  \centering
  \includegraphics[width=0.95\linewidth]{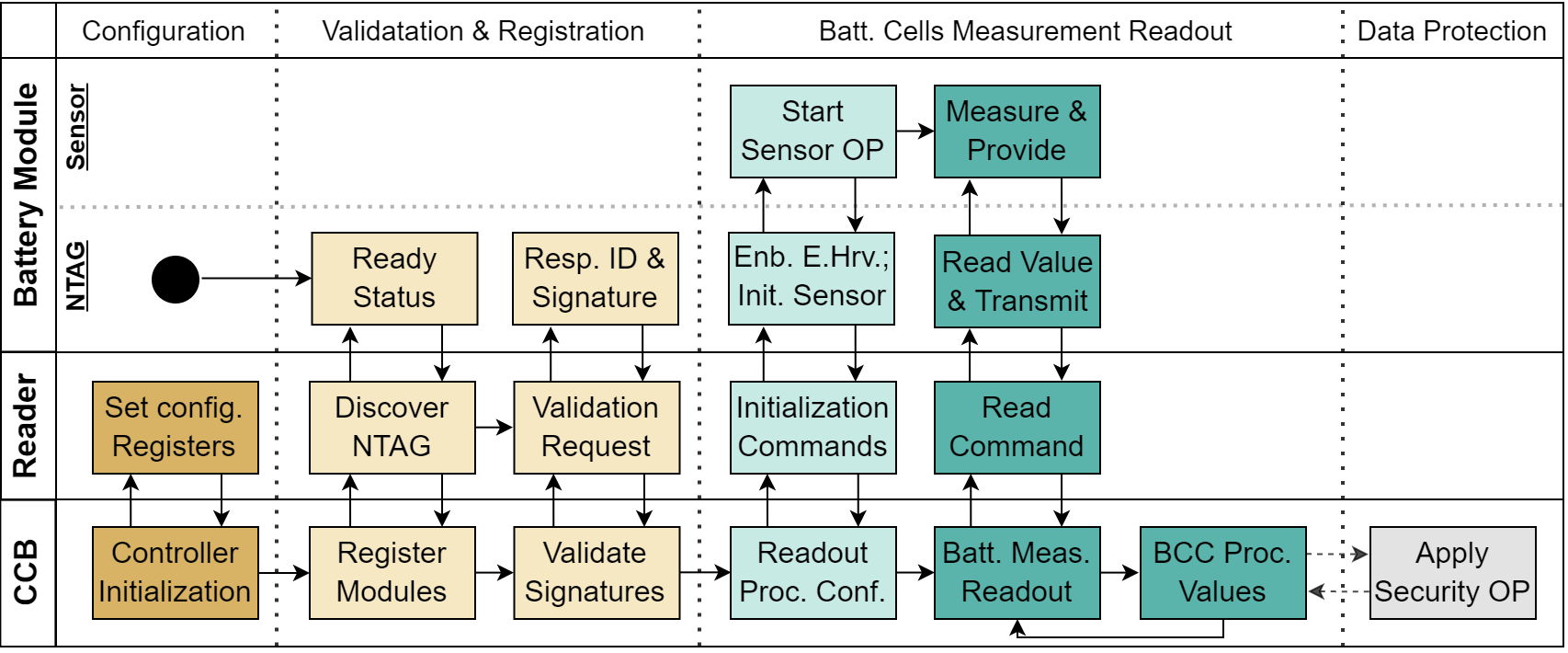}
  \caption{General system design representation using a swimlane sequence diagram showing the communication flow between CCB with its NFC Reader, and the battery module with NTAG and sensor. The process follows three main operational steps, with the security operations appliance being an optional step.}
  \label{fig:swimline_design}
\end{figure*}

\subsection{Energy Harvesting and Positioning}
\label{sec:design_eneg_harv}

A disadvantage that NFC has over most other wireless technologies is its relatively short range. This is of no issue in the presented design, as the BCCs and battery cells are usually tightly packed and installed together.
The NFC in the presented design uses the energy harvesting feature to power up the NTAG from the reader. The energy harvesting is also additionally used to power the necessary readout of the adjacent sensor. This feature limits the distance between the antennas. Depending on the environment, the distance peaks approximately at $5.4\,cm$. For a feasible communication and optimal initialization time, we opted to use a distance of $2\,cm$. 
The NTAG is not powered right at the boot-up of the system. It first needs to check if enough energy can be received from the present NFC field. The energy harvesting needs also to match the internally pre-configured voltage level. A voltage level of up to 3\,V can be supplied, which was also deemed sufficient for the sensor readout operation.
As both the sensor and the NTAG reside on the battery module, it would be possible for them to be directly powered as it is done in a conventional design. However, this characteristic is not present in our design model, as using the wiring to the battery modules would violate one of the design requirements set on reducing the extent of the necessary wires.

\subsection{Data Exchange Protocol}

The NFC reader is intended to establish the wireless connection to the dedicated battery sensors of the battery module. It plays the role of the active device, meaning that it initiates the communication. The sensors are able to transmit their values to the passive NTAG using the I2C connection. In this scenario, the master mode is used and the NTAG takes the role of the adapter module. The data is passed directly between the sensors and the NFC interface. Static Random Access Memory (SRAM) storage is used for the intermediate data placement before the read operation takes place. Additional commands had to be provided for the interaction on the battery module's I2C bus, as well as for the data transmission. These include the: (i) I2C read \& write commands, and (ii) content read; which allows direct content read from the intermediate SRAM storage. No MCU or any additional component is needed here, making the design relatively simple and cheap.

Fig~\ref{fig:swimline_design}. shows a swimlane sequence diagram that encapsulates all main operation processes intended to be covered during a common data exchange run. It covers the following steps:
\begin{enumerate}
    \item Configuration: initialization step at the session start, intended for loading up all the necessary configuration and operational material. It is expected to be run only once, usually at a start-up of a system (e.g., start of a vehicle). However, certain options could be cached, and hence pre-configured, with the aim of reducing the overall process execution time.
    \item Validation \& registration: CCB instructs its NFC reader to find and assign NTAGs first by using a discovery loop. Afterwards, validation takes place using the proposed signature authentication algorithm described more in detail in the Section~\ref{sec:secure_mech_auth}.
    \item Battery cells measurement readout: starts with the initialization step aimed for the measurement configuration. During this one-time procedure, the  NTAG initializes its communication with the sensors, but also enables the energy harvesting feature covered in Section~\ref{sec:design_eneg_harv}. After the initialization is finished, a process loop is run that, based on the sampling time, periodically reads out and processes the battery cell measurement data. The cells' data are further covered by the conventional BCC monitoring and diagnostic operations.
    \item Data protection: an optional step for the purpose of securing the read measurement, and BCC-derived, data. These operations are discussed in Sections \ref{sec:secure_mech_diag_prot} \& \ref{sec:secure_mech_sec_proc}. If used, it is intended to be included together with the measurement process loop. 
\end{enumerate}

%% file: security_mech.tex
\subsection{Battery Module Authentication Protocol}
\label{sec:secure_mech_auth}
In terms of security, NFC's advantage over the use of other wireless technologies is in both its short range and frequency band. This property limits the list of technologies that a potential attacker could use to attack the system. Since battery modules are usually enclosed in a protective case together with a BCC, the main potential attack vectors on these modules would be the ones initiated through counterfeiting~\cite{Lopez2017}. It is important that only battery cells that come from valid and approved manufactures are installed, as inadequate battery cells could potentially lead to hazards through compromising the BMS controller, or even going higher to the high-speed network outside the BMS environment \cite{cyber_sec_book}.

\begin{figure}[htbp]
  \centering
  \includegraphics[width=0.60\linewidth]{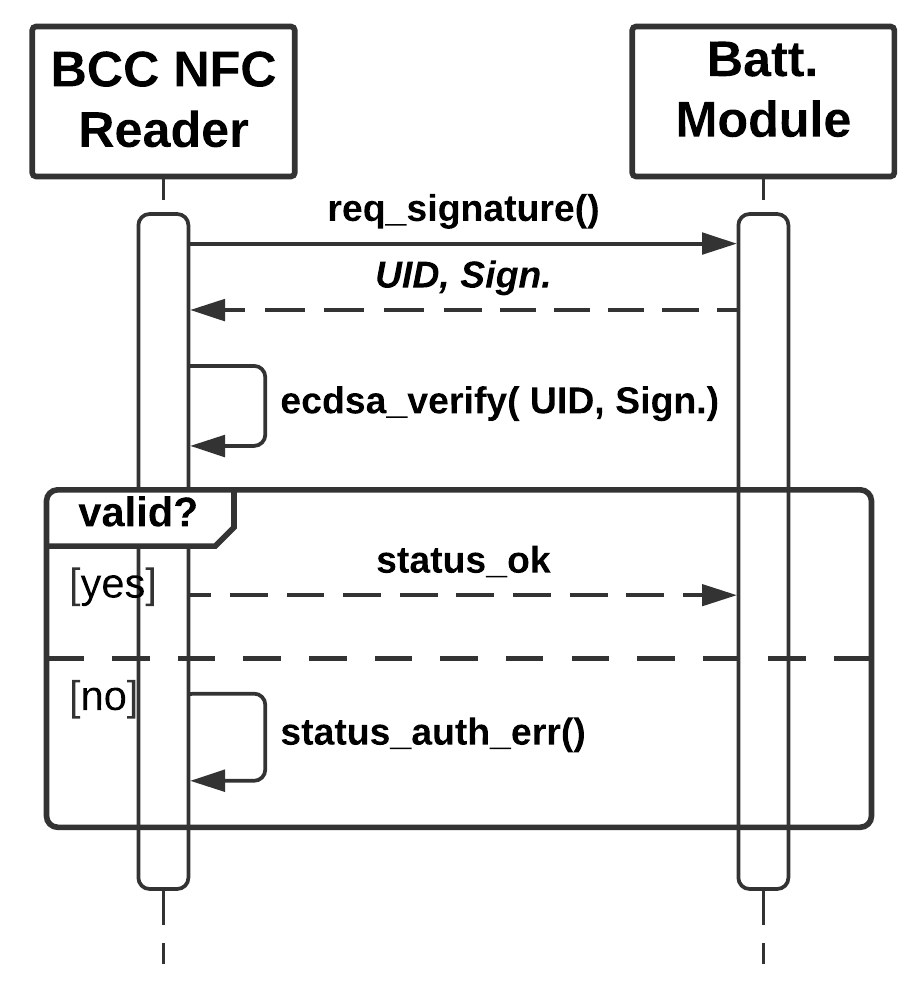}
  \caption{Sequence diagram of the authentication protocol.}
  \label{fig:bms_nfc_sequence_diag}
\end{figure}

To be able to securely verify that the battery modules are valid, we integrate the use of an authentication protocol in our design. This process is achieved by verifying a value that needs to be unique to each device.
Since NTAGs are usually shipped with a Unique Identifier (UID) value, we can use it as an input for an Elliptic Curve Digital Signature Algorithm (ECDSA). In our design, we use the \textit{secp128r1} protocol as the Elliptic Curve (EC) function, having a good balance between the performance and output sizes. The signature value, which is calculated with a private key during the manufacturing process or subsequently updated, is then stored in a protected memory space located on the NTAG chip. The BCC needs to have access to the public key, either it being pre-embedded or accessed through other secure channels. The authentication protocol is shown in Fig.~\ref{fig:bms_nfc_sequence_diag}. Before the signature verification takes place, the UID validity is first checked against the list of valid devices. Failure in either can lead to a warning message presented through the BMC controller, or a complete shutdown of the system, depending on the targeted use-case.

\subsection{BMS Status Data Protection}
\label{sec:secure_mech_diag_prot}
To protect the transmitted battery sensor data and the derived diagnostic data, it is necessary to apply different security measures. These measures would present an answer to the aforementioned security requirements and would be handled as an extension to the current BMS communication design, but also to its data acquisition protocol. Primarily, for the BMS use-case, it is important to fulfill the integrity and availability security requirements, since changes in the accuracy of the data and its sampling rate directly affects the output of the BMS control decisions. Data confidentiality also plays an important role, since the exposure of BMS data to unwanted third parties can also lead to the exposure of users' privacy, e.g., driver's behavior in electric vehicles. 

\begin{figure}[!ht]
  \centering
  \includegraphics[width=0.92\linewidth]{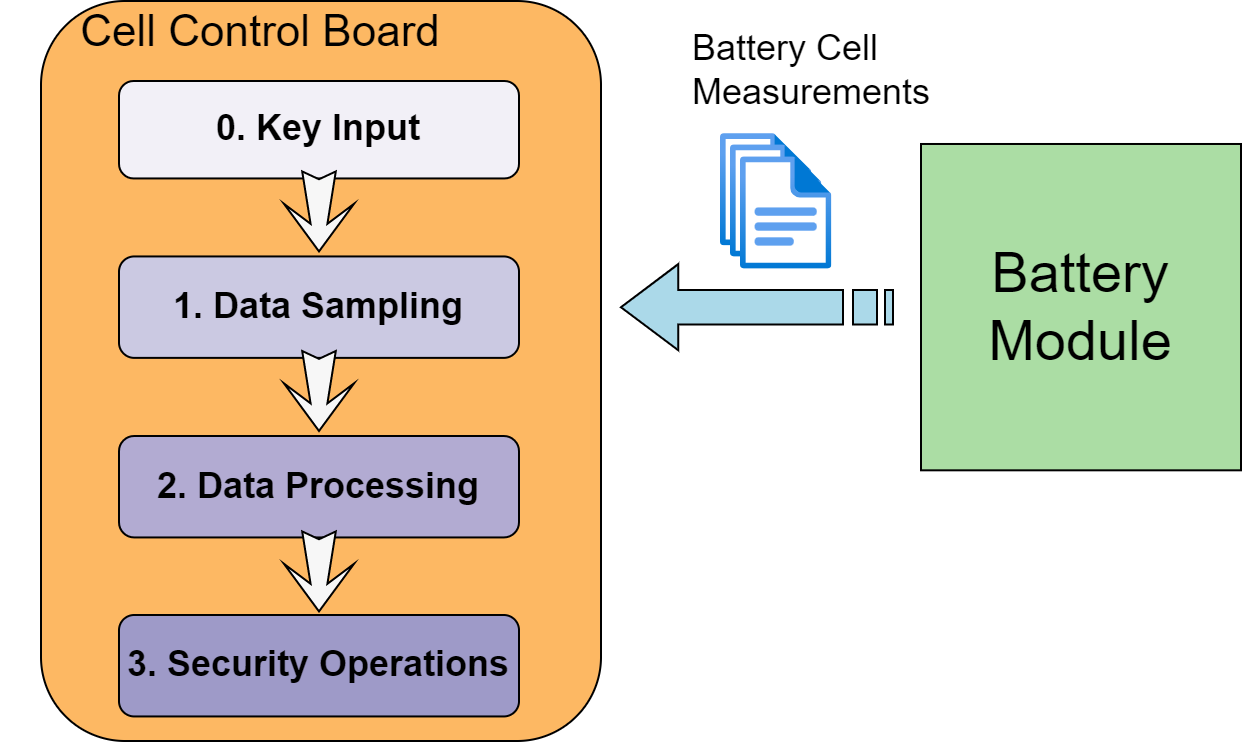}
  \caption{BMS measurement data sampling and secure handling.}
  \label{fig:data_sec_comm}
\end{figure}

Based on the design from Section~\ref{sec:design} and Fig.~\ref{fig:bms_design_architecture}, an extension in the view of a security module would be necessary as part of the CCB. This would free the design space of the battery module from the otherwise additional hardware modifications. It also means, however, that the transferred sensor data is not going to be encrypted or otherwise secured on the analogue connection between the battery module and the CCB, i.e., either through the proposed NFC interface or an adequate wired transfer. This is deemed to be acceptable, as the CCB and the battery module are usually tightly coupled and enclosed together, and attacks on those connections from the outside would be either difficult or even unfeasible. What is therefore important before the data transfer takes place between these modules, is that the authentication of the battery module was successful as described in Section~\ref{sec:secure_mech_auth}. Fig.~\ref{fig:data_sec_comm} shows the additional operational steps for secure data handling. The input of the key would take place at the start of the measurement session, and would be run only once for that session. Data sampling would contain the main functionality for receiving and applying monitoring and diagnostic operations from the standard BCC. Before the security operations can be applied, the data will first need to be structurally prepared, e.g., by using compression, or padding, during the data processing step. Finally, the designated security operations are run.

Placing the security operations on the CCB rather than onto the main BMS controller adds several benefits. Mainly, it presents an additional layer of security to otherwise different and uniform communication interfaces and standards used for the communication between the CCBs and the BMS controllers. It also frees the resources from the main BMS controller which would be necessary for secure storage in case of lifetime logging operations. Such data could then be stored onto the memory units connected to individual CCBs, encrypted and integrity protected against malicious modifications. This is especially important under the modulated topology where one BMS controller can communicate with multiple CCBs and would therefore reduce the computational and storage constraints on the main BMS controller. The CCB's controller needs to contain the necessary hardware and software components for the targeted security protocols.

\subsection{Security Protocols}
\label{sec:secure_mech_sec_proc}
To protect data confidentiality, it would be necessary to employ encryption of the sampled sensor data. Embedded devices rely on the use of either Hardware Security Modules (HSM), Secure Elements (SE), Trusted Platform Modules (TPM), or processor extensions with security function implementations. Security modules under the BMS use-case should be able to provide encryption and decryption operations, and tag verification for integrity check. The security module also provides other security functions, like a Random Number Generator (RNG), secure boot, and secure key generation and storage among others. The integrated algorithms are also often hardware-implemented, meaning that they benefit from the accelerated operations and physical security considerations.

Advanced Encryption Standard (AES) is often employed for symmetric encryption operations due to its high-security profile and small footprint. AES also benefits from hardware implementations for a faster algorithm execution. During employment, AES would need to be in different modes to provide encryption operation across a larger set of data. Traditionally, CBC and CTR modes are used, with Authenticated Encryption with Associated Data (AEAD) also gaining prominence where available, with modes like EAX, GCM, or CCM.

To protect the data against modifications, i.e., to guarantee its integrity, it is recommended to apply Message Authentication Code (MAC) calculations. These can be done on an arbitrary length of sampled sensor or diagnostic data either before or after the encryption on them took place. The calculated MAC bytes would be used for the integrity check. The MAC calculation can be left out in case the affiliated encryption algorithm is from the AEAD group and hence includes an integrity tag check as part of its procedure. 
These functions are sufficient in providing necessary BMS sensor data protection intended for either its intermediate storage or further data propagation and processing.

%% file: evaluation.tex
\begin{figure}[!t]
  \centering
  \includegraphics[width=0.95\linewidth]{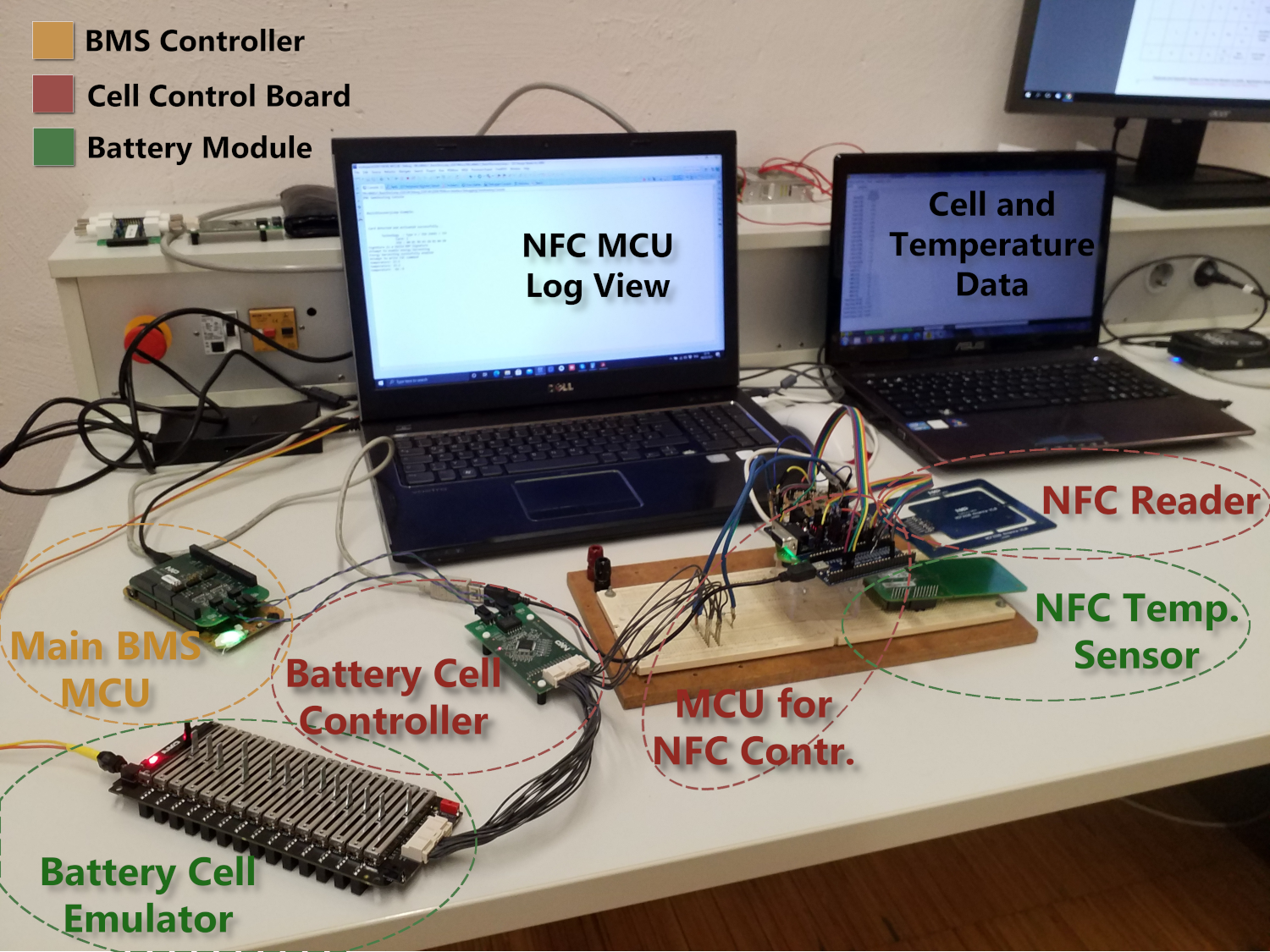}
  \caption{Evaluation setup for the BMS NFC sensor readout.}
  \label{fig:photo_test_setup}
\end{figure}

\subsection{Test System Implementation}
\label{eval:impl}

To test the presented design model, we implemented a test suite that contains the necessary BMS modules, as well as the additional NFC equipment. We aimed to use the NFC modules, which support the latest \textit{NFC Type 5 Tag} technology. Furthermore, the used components are automotive-graded where applicable for the purpose of replicating a real-world use-case as closely as possible. To that end, all devices used, except for the temperature sensor, come from the NXP Semiconductors lineup of products. The system is shown in Fig.~\ref{fig:photo_test_setup}.

As the main BMS controller, we use an S32K144 MCU board. It communicates with the CCB via the FRDMDUAL33664 shield over the TPL protocol. It is further connected to an RD33771CDST that houses an MC33771C, which functions as a BCC. The CCB contains an automotive NFC Reader for handling the NFC transmissions and another S32K144 as the MCU for programming and testing. The MCU board is connected with the NFC reader via SPI. The battery module consists of a BATT-14CEMULATOR that serves as a battery emulator, an NTAG component as the passive NFC device, and a BMP180 temperature sensor. The NFC devices are of the NCF33xx product family. The antennas of the active NFC reader and the passive NTAG devices are placed in parallel to each other, with the reader placed at a short distance over the NTAG, corresponding to the positioning discussion in Section~\ref{sec:design_eneg_harv}. The sensor is placed in close proximity with the NTAG device. 
For the setup, the temperature sensor from the battery emulator was disabled from transferring the temperature data, being otherwise routed through the attached NTAG component and the added BMP180 temperature sensor that communicates with the NTAG via the I2C protocol. Hence, the BCC is able to receive the emulated cell voltage data from the battery emulator, while the temperature sensor data is sent through the NFC interface. Both the temperature and the cell voltage data are first received by the BCC and then transmitted to the BMS controller. For the authentication protocol, we base our implementation on the \textit{originality signature} feature found on the NXP's RFID devices. Signature calculation and verification are handled via the \textit{ecc-nano} library \cite{eccnano}. Elements of the project development and evaluation were handled in a recent master's thesis~\cite{mastersthesis}.

\begin{figure}[!t]
  \centering
  \includegraphics[width=0.95\linewidth]{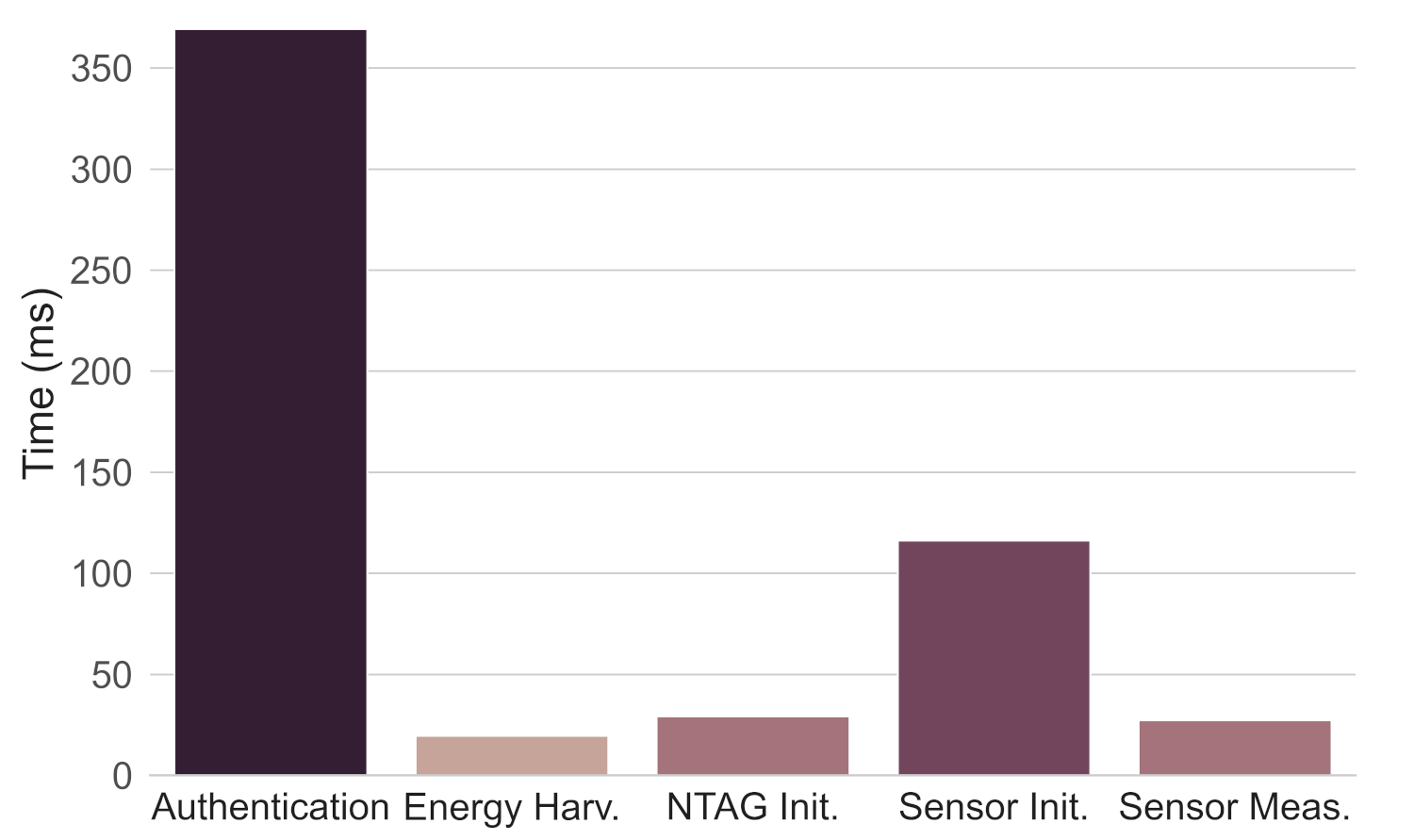}
  \caption{Time measurement results for the Initialization phase (Authentication, Energy Harv., NTAG Init., Sensor Init.) and Monitoring phase (Sensor Meas.).}
  \label{fig:bms_time_measurements}
\end{figure}

BMS status data protection: for this investigation, a security module was used to provide the necessary security operations, that comes integrated with the S32K144.  
The offered functionalities of this module are based on the Secure Hardware Extension (SHE) specification \cite{autosar}, and they included among others: a secure key derivation and storage, provided True Random Number Generator (TRNG), AES encryption algorithm with CBC mode, and Cipher-based MAC (CMAC) for data integrity and authentication. 

\subsection{NFC Sensor Readout Process Time Measurements}
\label{sec:time_meas}

We divide the main BMS monitoring process into two phases: (i) \textit{Initialization phase:} executed only once for device preparation and configuration, and (ii) \textit{Monitoring phase:} continuous action that is called on every sample step to measure and retrieve cell sensor data. Individual steps, as well as their time measurements, are shown in Fig.~\ref{fig:bms_time_measurements}. All represented time values are median values taken after multiple measurements.

The process starts after the NTAGs have already been discovered. As the first step the authentication protocol is run. This protocol run includes both sending an authentication request from the NFC reader, the response from the NTAG, and the verification calculation on the MCU that is connected with the NFC reader. The authentication step showed a median time of $369.30\pm0.37\,ms$, with majority of it being spent on the verification process. The relatively high execution time is attributed to this step being very hardware and software dependant, with optimizations being possible by using dedicated security components. With the NTAG verified, the energy harvesting check is handled which lasts for $19.64\pm0.25\,ms$. Finally, the NTAG operation initialization is run which measured $29.16\pm2.44\,ms$, followed with the sensor initialization that took $116.1\pm1.19\,ms$. After the initialization phase is finished, there is no need to reconfigure the devices during the system run. For the monitoring phase, sensor measurements are read and transmitted to the BCC using NFC communication. This phase is repeatable, with each action showing a time of $27.2\pm0.54\,ms$. 

\begin{figure}[htbp]
  \centering
  \includegraphics[width=0.92\linewidth]{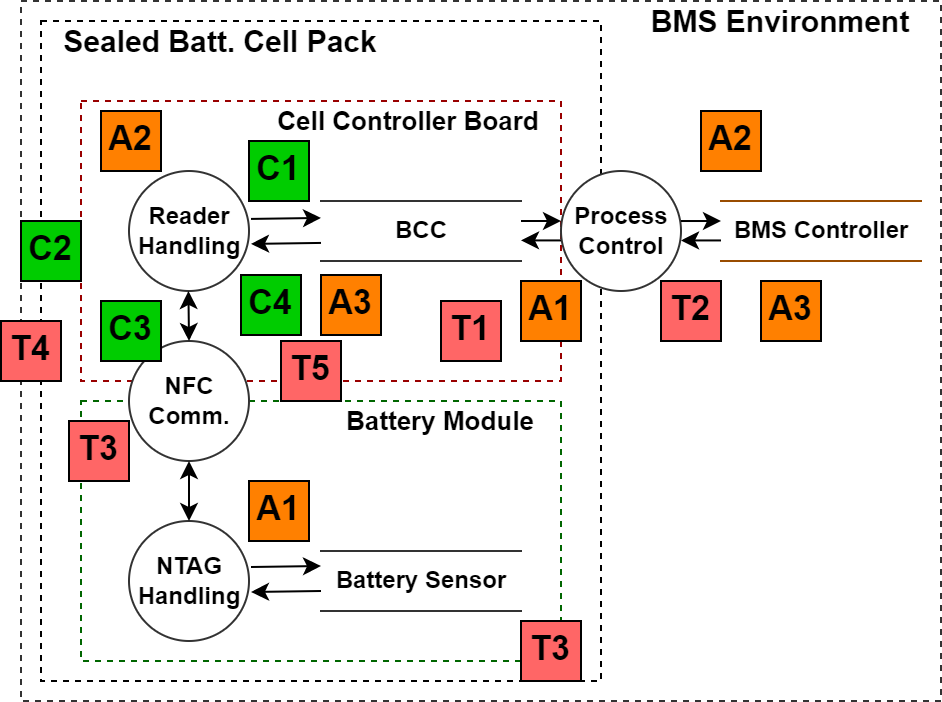}
  \caption{Security threat analysis visual overview using Data Flow Diagram.}
  \label{fig:thread_model}
\end{figure}

\subsection{Security Threat Analysis}
The proposed design has been subjected to a security threat analysis, for the purpose of evaluating the achieved security protection~\cite{Myagmar2005}. This has been conducted by listing individual \textit{Assets} (A), \textit{Threats} (T), and \textit{Countermeasures} (C). To better illustrate the carried out process, a visual representation of the targeted  use-case system model was made using Data Flow Diagram (DFD) which can be seen in Fig.~\ref{fig:thread_model}. Here, a demonstration is made with the indicated threats, their influenced assets, and answered countermeasures, also illustrating their potential points of impact. Threats are derived based on the carried security requirements analysis, as well as the basis security threats found in common BMS models done in prior research works \cite{Cheah2019, Sripad2017, Kumbhar2018}.

In our security model, we argue the following assumptions: (i) a battery module can only be communicated with via an adequate BCC, (ii) both the CCB and the battery module are enclosed in a chassis and the external communication can only be achieved through the BMS controller, (iii) every newly added and unknown battery module is considered untrustworthy, (iv) the CCB is deemed to contain adequate hardware and software components for security protection and calculations.

We indicate three important assets that need to be protected:
\begin{itemize}
    \item (A1)~\textit{Sensor data}: data retrieved from the cell sensors.
    \item (A2)~\textit{System integrity}: hardware and software integrity.
    \item (A3)~\textit{Diagnostic data}: status data derived from the monitored battery readings.
\end{itemize}

An attacker would look to exploit a vulnerability of the system, i.e., the potential to conduct a successful attack. Each attack is tied to a threat and assets that are targeted by it. 
In the following, each separate threat is listed with a given short description, the assets that it impacts, and the countermeasures:
\begin{itemize}
    \item \textbf{(T1)}~\textit{Battery control obstruction} $\mapsto (A1), (A3)$\\ 
    A potential threat that disturbs the cell balancing control through a fake source of sensor and diagnostic data.\\
    Mitigated through \textbf{(C1)}~\textit{Authentication through signature validation} by the proposed design. Here, BCCs validate every individual battery module, ensuring that the BMS controller only receives authorized status messages.
    \item \textbf{(T2)}~\textit{Tamper with BMS status messages} $\mapsto (A2), (A3)$\\ 
    A similar threat like (T1), but that is more covered and tries to tamper with the data rather than obstruct it.\\
    Also mitigated via the \textbf{(C1)} countermeasure. 
    \item \textbf{(T3)}~\textit{Backdoor access} $\mapsto (A1), (A2)$\\ 
    An attacker might try to gain system's access through either the NFC interface or a counterfeited battery module.\\
    Protected through the \textbf{(C1)} countermeasure, but also by reducing the attack proximity by relying on the \textbf{(C3)}~\textit{NFC physical layer characteristics}.
    \item \textbf{(T4)}~\textit{Remote attack} $\mapsto (A1), (A2), (A3)$\\ 
    Various attacks can be launched from outside of the system on unprotected channels by using wireless communication. Under this context, we primarily consider the probing attacks that target the NFC channels.\\
    \textbf{(C2)} \textit{Cell pack sealing} protects against remote attacks by isolating interfaces via material shielding. Also, \textbf{(C3)} would hamper the possibility of such an attack through frequency spectrum and range limitations.
    \item \textbf{(T5)} \textit{BMS log data compromise} $\mapsto (A1), (A3)$\\ 
    Such an attack can take place on CCB, both from the local or possible backdoor access via exposed (T3), or through (T4). These include both the privacy leak of the associated system through the compromise of the read raw data, but also any kind of unauthorized data changes which would be intermediately stored on the CCB.\\
    The data can be protected by \textbf{(C5)}~\textit{Data security measures} which include the prescribed encryption, authentication and integrity validations. 
\end{itemize}

\begin{table}[htp] \centering
\caption{ Full BMS Data Sampling, Processing and Security Operation }
\label{table:full_times_data_process}
\begin{tabular}{@{\extracolsep{1pt}} lccc}
\\[-1.8ex]\hline 
\hline \\[-1.8ex] 
\textbf{Iterations} & \multicolumn{1}{C{1.45cm}}{1 sample} & \multicolumn{1}{C{1.45cm}}{5 samples} & \multicolumn{1}{C{1.45cm}}{100 samples} \\
\hline \\[-1.8ex] 
\textbf{Time} & $114.85\pm0.73\,ms$ & $580.56\pm1.55\,ms$ & $11.64\pm0.02\,s$ \\
\hline \\[-1.8ex] 
\end{tabular}
\end{table}

\subsection{Data Security Overhead Analysis}
\label{sec:sec_overhead_analysis}

An evaluation was conducted for the purpose of testing the BMS data security handling. This evaluation includes a model that was built to depict a real-world representation of the BMS data structure that includes both the monitoring and diagnostic data components. The evaluation follows the design principles described in Section~\ref{sec:secure_mech_diag_prot} and security protocol considerations in Section~\ref{sec:secure_mech_sec_proc}. To fulfill the security conditions, we employ the use of a security module as stated in Section~\ref{eval:impl}.

The BMS test system uses a battery emulator that emulates 14 cell voltages together with a sensor temperature value derived from the extended NFC measurement components. The software presents each measurement with an identifier and the measured value. These values are considered monitoring values. The BCC is further capable of deriving diagnostic values for the active status report. One-time reading from one battery module is considered a sample. In our testing case, one such sample has a length of 162 bytes. For security purposes, padding is added to round up the total size to be 176 bytes, a multiple value of 16,
since the security algorithms used are of the 128-bit block length.

The evaluation was divided into three phases following the design given in Fig.~\ref{fig:data_sec_comm}. Values are shown as mean values derived from multiple measurements. The initial key insertion step was measured at a constant $20\,ms$. \raisebox{.5pt}{\textcircled{\raisebox{-0.9pt} {1}}} \textit{Data sampling} was measured at $112.98\pm0.54\,ms$. Among this, the measurement step only required $3.5\,ms$, with the remaining $109.5\,ms$ being used for the diagnostic derivations. \raisebox{.5pt}{\textcircled{\raisebox{-0.9pt} {2}}} \textit{Data processing}  was shown to have little impact and be very fast with a resulting time of $1.0\pm0.1\,ms$.
\raisebox{.5pt}{\textcircled{\raisebox{-0.9pt} {3}}} \textit{Security operations} include the AES-CBC and CMAC calculations for the data confidentiality, integrity and authenticity security coverage. The execution was relatively fast, resulting in a total time of $992\pm8.75\,\mu s$.

As it can be concluded from the evaluation, the main operational overhead comes from the data sampling step. This shows that the integration of security operations along with the traditional BMS data sampling results in minimal overhead change, having an increase of $1.7\,\%$, and therefore would not intervene with the standard BMS time-critical safety operations. Even if the security data logging procedure would only be limited to the measurement steps, either due to performance of infrequent diagnostic checks, it still would result in an acceptable overhead range, adding additional $\approx1.5\,ms$ to the $3.5\,ms$ of measurements sampling.
The main challenges would come from determining how often the security logging should take place, and defining what would be the necessary memory capacity for the long-term administration. The measurements were also done over a longer operation run, but no significant changes between the measurements have been detected. The results follow a linear time increase. Total times for 1, 5 and 100 sample runs are shown in Table~\ref{table:full_times_data_process}.

\begin{figure*}[!ht]
  \centering
  \includegraphics[width=1.0\linewidth]{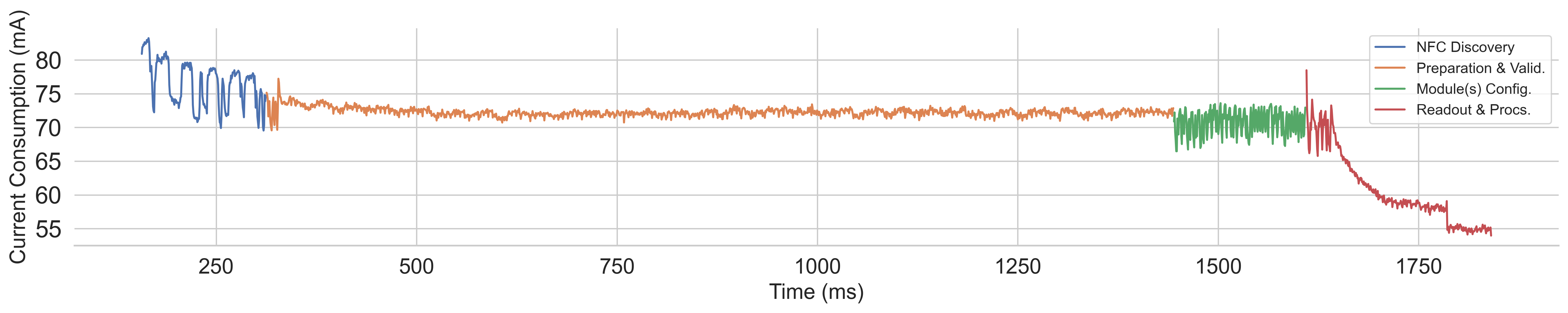}
  \caption{Current consumption over time for the process MCU in CCB during the start-up configuration period for the active battery sensor readout showcasing: discovery process from the NFC reader, internal config. \& secure NTAG validation, batt. module, sensor and NTAG configuration, and NFC readout process.}
  \label{fig:current_consump_mcu_reader}
\end{figure*}

\subsection{Energy Consumption}
\label{sec:eval_energy_cons}

We measured the energy consumption of our BMS implementation to investigate how much additional energy would be required with the added CCB components, energy overhead for the added sensors and the NTAG, and the overall energy consumption for the BMS controller when considering the added security operations. For conducting the measurements, we used a Nordic Semiconductor Power Profiler Kit.


\begin{figure}[!ht]
  \centering
  \includegraphics[width=1.0\linewidth]{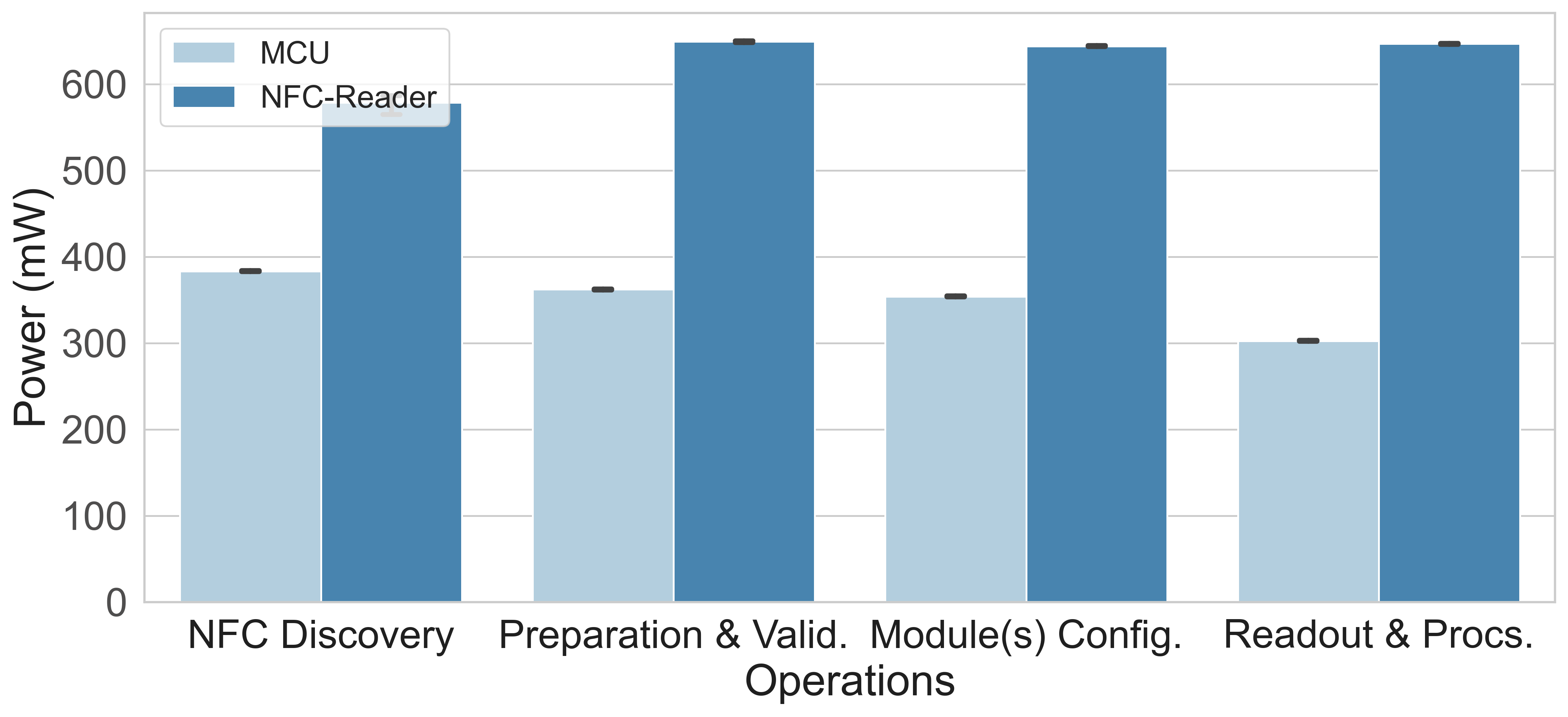}
  \caption{Power consumption over the CCB's MCU and NFC reader.}
  \label{fig:power_consumption_devices}
\end{figure}

The \textbf{CCB} was evaluated on two added elements: on the extended process MCU, and the active NFC reader. Fig.~\ref{fig:current_consump_mcu_reader} shows the current consumption for the CCB's MCU, which is responsible for the control of the NFC reader. The same operational segments were also considered in parallel when measuring the consumption from the CCB's NFC reader board.

From Fig.~\ref{fig:current_consump_mcu_reader}, we can observe four operational segments:
\begin{enumerate}
    \item \textit{NFC Discovery}: mainly considers the discovery loop for the battery module's passive NTAG component; shows the highest peak in current consumption, but the average remains consistent with other operations. 
    \item \textit{Preparation \& Validation}: board configuration and start-up steps; also includes the signature authentication step (Section~\ref{sec:secure_mech_auth}). Shows a constant and stable consumption over most of its period.
    \item \textit{Module(s) Configuration}: configuration command exchange for NFC and sensor devices on the battery module. More oscillating consumption due to a more intensive NFC reader interaction.
    \item \textit{Readout \& Processing}: one iteration of the battery sensor readout and data handling. The drop in current consumption indicates inactivity on the MCU part after the operation ends, wherein the beginning a higher consumption can be observed from the NFC data exchange.
\end{enumerate}

Fig.~\ref{fig:power_consumption_devices} shows the graphical comparison for average power consumption between the CCB's MCU and NFC reader after five different measurement runs. The operational voltage for both components was set at $5\,V$. Overall, the added devices resulted in an increase of up to $1\,W$ of power consumption, without optimization considerations. This means that for the repeatable monitoring phase (described in Section~\ref{sec:time_meas}), without any other additional computational overhead, the energy consumption amounts to $25.82\,mJ$.

The power consumption shown considers the consumption of the whole NFC reader board during the active period. This means that it accounts for all regulators, communication interfaces, the NFC chip controller, and most importantly, the active RF transceiver. Since for most of the operational run, the communication interaction between the active and passive NFC devices were taking place, the RF field remained also mainly active. In a general environment, NFC readers are indented to offer a polling feature, i.e., periodical wake-up from the stand-by state for the purpose of detecting present passive NFC devices. This feature greatly reduces the average current consumption over time. However, in the presented design, the communication remains active for most of the time during the monitoring phase since the positioning condition of the devices would not change and, depending on the sampling rate, the next measurement might occur soon after the last one finished. Optimisation and adjustment of the standby mode and RF activation is largely dependent on the targeted system implementation goals and is left open for the developers.

\textbf{Battery pack sensor} consumption was negligible compared to other energy consumption of the system, with current consumption of $40\, nA$ for the standby state, and peeking for a short time of up to $22\,uA$ during initialization and active period. 
The NTAG relies on the energy harvesting feature for the operation and control of the sensors (see Section~\ref{sec:design_eneg_harv}). In our test case, this results in additional current consumption draw of the NFC reader, with an average rise of $5\,mA$, when in the range of the NTAG.

We analyzed the \textbf{BMS controller} on the total power and energy consumption for one full diagnostic sampling cycle. The average drawn power resulted in $122.16\,mW$, with an average energy consumption of $13.80\,mJ$, after ten different system runs. Additionally, the security operations, and its preceding data processing, resulted only in a slight increase of energy consumption with $0.28\,mJ$ for one sampling cycle, i.e., $2.66\,mJ$ for one-time key-insertion operation. We can observe that the added security operations result only in a minimal increase of up to $2\,\%$ of energy consumption per sample run.

\subsection{Battery Sensor Throughput Analysis}

The throughput of battery module sensor data largely depends on several factors. Primarily, it is dependent on: (i) the number of the total sensors used per module, (ii) the number of total battery modules used per CCB, the number of CCBs used per the main central BMS controller, and in this case of using the NFC components, (iv) the total number of communicating NFC components (active and passive) and the number of sensors per passive components (communication chains). As indicated in Section~\ref{eval:impl}, for the experimental setup a battery emulator was used that offers the reading of fourteen battery cells and one temperature sensor. As such, under our setup, we represent a system that has: one sensor per battery module, one passive NTAG device per battery module, connected directly with the battery sensor, and a CCB with one active NFC reader per assigned battery module. More points on the realization and potential future work based on the aforementioned throughput factors are discussed in Section~\ref{sec:discussion}.

\begin{table}[!t] \centering
\caption{Expected Sampling Throughput Per One CCB and Batt. Module} 
\label{table:full_time_battery_throughput}
\begin{tabular}{@{\extracolsep{1.5pt}} lccc}
\\[-1.8ex]\hline 
\hline \\[-1.8ex] 
\multicolumn{1}{C{1.35cm}}{\textbf{Iterations}} & \multicolumn{1}{C{1.45cm}}{100} & \multicolumn{1}{C{1.45cm}}{1,000 } & \multicolumn{1}{C{1.45cm}}{10,000} \\
\hline \\[-1.8ex] 
\multicolumn{1}{C{1.35cm}}{\textbf{Time \& Data Size}} & \multicolumn{1}{C{1.45cm}}{$2.96\,s$\;\; $0.76\,kB$}  & \multicolumn{1}{C{1.45cm}}{$29.24\,s$ $7.54\,kB$} & \multicolumn{1}{C{1.45cm}}{$297.45\,s$ $76.68\,kB$} \\
\hline \\[-1.8ex] 
\end{tabular}
\end{table}

The readout of the NTAG is done through the provided SRAM. The SRAM in our test environment offers 256 bytes of data transfer, with data being divided into blocks of 4 bytes. In our setup, reading the whole SRAM would take $82.28\,ms$. However, in a real setting, this readout would probably require much less data. As indicated, each sensor would need 1-2 blocks containing 4 bytes each to process and send its derived data. The amount of sensors is also usually limited per pack, and it is very unlikely, that with current battery modules the data requirements would exceed one SRAM read request. 


Each measurement requires three actions to take place:
\begin{enumerate}
\item CCB (NFC Read.) $\rightarrow$ (NTAG) Batt. module; write command to enable and start the sensor measurement.
\item CCB (NFC Read.) $\rightarrow$ (NTAG) Batt. module; read command to read out from the specific block of the SRAM of the saved sensor values.
\item CCB (NFC Read.) $\leftarrow$  (NTAG) Batt. module; transmitting the sensor values from the NTAG's SRAM.
\end{enumerate}

The request and response frames contain additional data in the form of flags, IDs, commands, address, and the Cycle-Redundancy-Check (CRC) appendices. Thus, the first two write and read SRAM commands take additional 15 bytes of the header, which can be reduced to 7 bytes if the ID component is removed (if the communication is 1-to-1, it is not necessary). The response SRAM read frame only has 3 header bytes (flags and CRC). The remaining payload depends on the sensor data, which in our case is two blocks, i.e., 8 bytes. Out of those, the measured value is contained in 2 bytes.

Based on the experiments from the implementation setup, the CCB was able to conduct 33 measurements per second, i.e, for reading 8 bytes, 264 bytes/s of the pure measurement data. As noted, each measurement is a three-stage process with always the same repeating overhead. The time required for processing the received data by the CCB is negligible compared to the transfer time. This accounted for $1-2\,ms$ between each read request used for handling the processing of the received sensor data, but measurements otherwise correspond to running a single sensor measurement during the ``Monitoring phase'' as indicated in Section~\ref{sec:time_meas}. The total amount of time and handled sensor payload data when running repeatable measurements for a different number of iterations is shown in Table~\ref{table:full_time_battery_throughput}. As expected, linear growth of time compared to the number of repeated iterations is observed.

Compared to the measurements conducted for the overall BMS process after applying security operations in Section~\ref{sec:sec_overhead_analysis}, it can be concluded that the amount of data processed would suffice for the current setup, even when using multiple sensors per battery module. Furthermore, for a modulated topology that is presented here, the measurements and sampling would be conducted in parallel, and are independent of the number of used CCB, being only limited by the processing power of the assigned BMS controller. However, for BMS that have requirements for a faster sampling rate, in this case, that being $<30\,ms$, additional optimization aspects would need to be considered.

%% file: discussion_future_work.tex
As we see from Section~\ref{sec:time_meas}, the initialization phase is very time-demanding. The primary reason for this is the long execution time for the signature validation, which took around 69\% of the total initialization phase time in case all four initialization phase steps would be executed sequentially. Based on our evaluation tests and findings, we note the following important points that can be handled during the implementation of the proposed design to alleviate the time and help in phase delivery: (a) signature validation hardware and software need to be optimized for the target system to reduce the overall initial execution time, (b) process parallelization for steps reduction in the execution, (c) configuration and status caching for the targeted devices.

The proposed system design solution can also be used on different BMS settings regardless of the use case, which should fit the needs of automobile and industrial environments. In this context, the applicability of the presented solution is not only limited to conventional temperature sensors but also other sensors as part of a battery module. The battery cell's sensor placement and the target of measurement play an important role and could benefit from using the means of NFC transfer. The closer the sensor is to the core of the battery, the more accurate and time-punctual results are going to be. To this end, it would be possible to utilize NFC to transfer the data from the inside to the outside of the battery from these sensors. These measurements would add an additional layer to the safety precautions of the BMS, and hence, would influence the increased safety of the overall system. Separate research would need to be conducted which would investigate the optimal placement and usability of using the NFC communication for the data transfer in regards to the actual sensor placement in a battery module.

Next to the handling of the sensor, antenna positioning should be further investigated as well \cite{6966212}. For the current setup, a parallel placement is used with no physical considerations. Future work should also include research on the limits of the NFC range when considering the obstructing environment of the enclosed BMS modules. Additionally, an analysis should be made on the possible range and performance when not considering the energy harvesting feature. As mentioned in this work under Section~\ref{sec:design_eneg_harv}, the proposed design uses the energy harvesting feature of the NFC technology to allow for less reliance on the wired connections with the batteries. However, by disabling this feature and basing the use-case on using the source of power from the underline batteries, the range can be greatly exceeded, but at a higher cost, as additional wiring would also need to be provided.

Concerning the system design, another important point to consider for future work is the analysis of the number of used NFC elements. The current design proposes the use of one active NFC device per CCB and one passive device for the battery module. Considerations should be made on the adequate distribution of active and passive NFC devices. This is especially important for the passive NFC devices, i.e., the NTAGs, since an adequate hardware solution should be provided that considers the potential of multiple sensors placed in one battery module, or multiple NTAGs being handled by one active NFC component. Optimisations in the system design could lead to a reduction in the overall production cost.

In this section, we have primarily discussed the hardware aspects of future work and improvement, but an investigation should also be made into the optimization methods for the purpose of improving the time execution and reliability of the connection during the NFC sensor readout process. This can be realized on different software layers, targeting both the lower driver control and application stack. Among others, this investigation may include the consideration of different communication protocol extensions, but also the improved security realization. For future work, we also plan to further extend the investigation of the data security control within the BMS environment. Security attention should be given to the extension of the authentication algorithm, but also in adding an extra security layer for communication with the external components and services. Additional threat aspects need to be considered when the attack surface is extended \cite{cyber_sec_book, Cheah2019}.

%% file: conclusion.tex
In this work, we have presented the idea of using NFC as a wireless communication interface for battery sensor readouts in BMS.
A system design has been proposed that considers the construction of a modulated BMS with NFC components with special regard to the data exchange protocol and NFC requirements. To alleviate the risk of the counterfeited battery cells and prevent safety and security threats that can arise from them, an authentication model has been proposed and evaluated. A further study has been conducted that investigates the security handling and control of the derived sensor and diagnostic data once they are logged on a cell control board. Experimental results using real components show the feasibility of our approach, but also design challenges that open the possibilities of various further research in this field.

%% file: main.bbl
\begin{thebibliography}{10}

\bibitem{Hu2019}
X.~Hu, F.~Feng, K.~Liu, L.~Zhang, J.~Xie, and B.~Liu, ``{State estimation for
  advanced battery management: Key challenges and future trends},'' {\em
  Renewable and Sustainable Energy Reviews}, vol.~114, 2019.

\bibitem{8168251}
R.~Xiong, J.~Cao, Q.~Yu, H.~He, and F.~Sun, ``{Critical Review on the Battery
  State of Charge Estimation Methods for Electric Vehicles},'' {\em IEEE
  Access}, vol.~6, pp.~1832--1843, 2018.

\bibitem{6532486}
H.~Rahimi-Eichi, U.~Ojha, F.~Baronti, and M.-Y. Chow, ``{Battery Management
  System: An Overview of Its Application in the Smart Grid and Electric
  Vehicles},'' {\em IEEE Industrial Electronics Magazine}, vol.~7, 2013.

\bibitem{andrea2010battery}
D.~Andrea, {\em {Battery Management Systems for Large Lithium-ion Battery
  Packs}}.
\newblock EBL-Schweitzer, Artech House, 2010.

\bibitem{fireBattery}
P.~Sun, R.~Bisschop, H.~Niu, and X.~Huang, ``{A Review of Battery Fires in
  Electric Vehicles},'' {\em Fire Technology}, pp.~1--50, 01 2020.

\bibitem{7357002}
G.~De~Maso-Gentile, A.~Bacà, L.~Ambrosini, S.~Orcioni, and M.~Conti, ``{Design
  of CAN to Bluetooth gateway for a Battery Management System},'' in {\em 2015
  12th International Workshop on Intelligent Solutions in Embedded Systems
  (WISES)}, pp.~171--175, 2015.

\bibitem{7151581}
C.~Shell, J.~Henderson, H.~Verra, and J.~Dyer, ``{Implementation of a Wireless
  Battery Management System (WBMS)},'' in {\em 2015 IEEE International
  Instrumentation and Measurement Technology Conference (I2MTC) Proceedings},
  pp.~1954--1959, 2015.

\bibitem{Rahman_2017}
A.~Rahman, M.~Rahman, and M.~Rashid, ``{Wireless Battery Management System of
  Electric Transport},'' {\em {IOP} Conference Series: Materials Science and
  Engineering}, vol.~260, p.~012029, nov 2017.

\bibitem{electronics10182193}
A.~Samanta and S.~S. Williamson, ``{A Survey of Wireless Battery Management
  System: Topology, Emerging Trends, and Challenges},'' {\em Electronics},
  vol.~10, no.~18, 2021.

\bibitem{6229439}
M.~Schneider, S.~Ilgin, N.~Jegenhorst, R.~Kube, S.~Püttjer, K.-R.
  Riemschneider, and J.~Vollmer, ``{Automotive Battery Monitoring by Wireless
  Cell Sensors},'' in {\em 2012 IEEE International Instrumentation and
  Measurement Technology Conference Proceedings}, pp.~816--820, 2012.

\bibitem{7385582}
D.~Alonso, O.~Opalko, and K.~Dostert, ``{Physical Layer Performance Analysis of
  a Wireless Data Transmission Approach for Automotive Lithium-Ion
  Batteries},'' in {\em IEEE VNC}, pp.~235--242, 2015.

\bibitem{6914889}
M.~Lee, J.~Lee, I.~Lee, J.~Lee, and A.~Chon, ``{Wireless Battery Management
  System},'' in {\em World Electric Vehicle Symposium and Exhibition (EVS27)},
  pp.~1--5, 2013.

\bibitem{wireless_battery_arpa}
J.~Farmer, J.~Chang, J.~Zumstein, J.~Kotovsky, E.~Zhang, A.~Dobley, G.~Moore,
  F.~Puglia, S.~Osswald, K.~Wolf, J.~Kaschmitter, S.~Eaves, and T.~Bandhauer,
  ``{Wireless Battery Management System for Safe High-Capacity Li-Ion Energy
  Storage},'' tech. rep., {Lawrence Livermore National Laboratory}, 01 2013.

\bibitem{wired_vs_wireless}
T.~Vogt, ``{Wired vs. Wireless Communications in EV Battery Management},''
  tech. rep., Texas Instruments, 10 2020.

\bibitem{8591253}
T.~Gherman, M.~Ricco, J.~Meng, R.~Teodorescu, and D.~Petreus, ``{Smart
  Integrated Charger with Wireless BMS for EVs},'' in {\em IECON - 44th Annual
  Conference of the IEEE Industrial Electronics Society}, 2018.

\bibitem{6966212}
D.~Alonso, O.~Opalko, M.~Sigle, and K.~Dostert, ``Towards a wireless battery
  management system: Evaluation of antennas and radio channel measurements
  inside a battery emulator,'' in {\em 2014 IEEE 80th Vehicular Technology
  Conference (VTC2014-Fall)}, pp.~1--5, 2014.

\bibitem{9090905}
K.~Lounis and M.~Zulkernine, ``{Attacks and Defenses in Short-Range Wireless
  Technologies for IoT},'' {\em IEEE Access}, vol.~8, 2020.

\bibitem{Sripad2017}
S.~Sripad, S.~Kulandaivel, V.~Pande, V.~Sekar, and V.~Viswanathan,
  ``{Vulnerabilities of Electric Vehicle Battery Packs to Cyberattacks},'' {\em
  ArXiv}, 2017.

\bibitem{Kumbhar2018}
S.~Kumbhar, T.~Faika, D.~Makwana, T.~Kim, and Y.~Lee, ``{Cybersecurity for
  Battery Management Systems in Cyber-Physical Environments},'' {\em ITEC
  2018}, pp.~761--766, 2018.

\bibitem{8813669}
A.~Khalid, A.~Sundararajan, A.~Hernandez, and A.~I. Sarwat, ``{FACTS Approach
  to Address Cybersecurity Issues in Electric Vehicle Battery Systems},'' in
  {\em 2019 IEEE TEMSCON}, pp.~1--6, 2019.

\bibitem{101093}
S.~Engels, ``{Counterfeiting and piracy: the industry perspective},'' {\em
  Journal of Intellectual Property Law \& Practice}, vol.~5, 05 2010.

\bibitem{9617320}
F.~Basic, M.~Gaertner, and C.~Steger, ``{Towards Trustworthy NFC-based Sensor
  Readout for Battery Packs in Battery Management Systems},'' in {\em 2021 IEEE
  International Conference on RFID Technology and Applications (RFID-TA)},
  pp.~285--288, 2021.

\bibitem{10.1145/3331052.3332471}
M.~Sp\"{o}rk, C.~A. Boano, and K.~R\"{o}mer, ``{Performance and Trade-Offs of
  the New PHY Modes of BLE 5},'' in {\em Proceedings of the ACM MobiHoc
  Workshop on Pervasive Systems in the IoT Era}, PERSIST-IoT '19, (New York,
  NY, USA), p.~7–12, Association for Computing Machinery, 2019.

\bibitem{9236279}
X.~Huang, A.~B. Acharya, J.~Meng, X.~Sui, D.-I. Stroe, and R.~Teodorescu,
  ``{Wireless Smart Battery Management System for Electric Vehicles},'' in {\em
  2020 IEEE Energy Conversion Congress and Exposition (ECCE)}, pp.~5620--5625,
  2020.

\bibitem{en11010125}
T.~Kim, D.~Makwana, A.~Adhikaree, J.~S. Vagdoda, and Y.~Lee, ``{Cloud-Based
  Battery Condition Monitoring and Fault Diagnosis Platform for Large-Scale
  Lithium-Ion Battery Energy Storage Systems},'' {\em Energies}, vol.~11,
  no.~1, 2018.

\bibitem{LI2020101557}
W.~Li, M.~Rentemeister, J.~Badeda, D.~Jöst, D.~Schulte, and D.~U. Sauer,
  ``{Digital twin for battery systems: Cloud battery management system with
  online state-of-charge and state-of-health estimation},'' {\em Journal of
  Energy Storage}, vol.~30, p.~101557, 2020.

\bibitem{Lopez2017}
A.~B. Lopez, K.~Vatanparvar, A.~P. Deb~Nath, S.~Yang, S.~Bhunia, and M.~A.
  Al~Faruque, ``{A Security Perspective on Battery Systems of the Internet of
  Things},'' {\em Journal of Hardware and Systems Security}, 2017.

\bibitem{ISO18092}
{ISO/IEC 18092:2013}, ``{Information Technology—Telecommunications and
  Information Exchange Between Systems—Near Field Communication—Interface
  and Protocol (NFCIP-1)},'' {Standard}, International Organization for
  Standardization, 2013.

\bibitem{ISO21481}
{ISO/IEC 21481:2021}, ``{Information technology — Telecommunications and
  information exchange between systems — Near field communication interface
  and protocol 2 (NFCIP-2)},'' {Standard}, International Organization for
  Standardization, 2021.

\bibitem{8098906}
T.~Ulz, T.~Pieber, C.~Steger, S.~Haas, and R.~Matischek, ``{Sneakernet on
  Wheels: Trustworthy NFC-based Robot to Machine Communication},'' in {\em 2017
  IEEE International Conference on RFID Technology Application (RFID-TA)},
  pp.~260--265, 2017.

\bibitem{6228335}
C.-L. Chen, Y.-Y. Chen, T.-F. Shih, and T.-M. Kuo, ``{An RFID Authentication
  and Anti-counterfeit Transaction Protocol},'' in {\em 2012 International
  Symposium on Computer, Consumer and Control}, pp.~419--422, 2012.

\bibitem{9075232}
B.~A. Alzahrani, K.~Mahmood, and S.~Kumari, ``{Lightweight Authentication
  Protocol for NFC Based Anti-Counterfeiting System in IoT Infrastructure},''
  {\em IEEE Access}, vol.~8, pp.~76357--76367, 2020.

\bibitem{cyber_sec_book}
N.~Druml, A.~Genser, A.~Krieg, M.~Menghin, and A.~Höller, {\em {Solutions for
  Cyber-Physical Systems Ubiquity}}.
\newblock IGI Global, 08 2017.

\bibitem{eccnano}
``{ECC-Nano}.'' https://github.com/iSECPartners/nano-ecc, 2013.
\newblock Accessed: 15.01.2022.

\bibitem{mastersthesis}
M.~Gaertner, ``{Design and Implementation of a NFC-based Solution for Secure
  Battery Management Systems},'' Master's thesis, Graz University of
  Technology, 2021.

\bibitem{autosar}
AUTOSAR, {\em {Specification of Secure Hardware Extensions}}, 2019.

\bibitem{Myagmar2005}
S.~Myagmar, A.~Lee~J., and W.~Yurcik, ``{Threat Modeling as a Basis for
  Security Requirements},'' {\em in SREIS}, 2005.

\bibitem{Cheah2019}
M.~Cheah and R.~Stoker, ``{Cybersecurity of Battery Management Systems},'' {\em
  HM TR series}, vol.~10, no.~3, p.~8, 2019.

\end{thebibliography}
